\documentclass[prb,superscriptaddress,showpacs,amsmath,amssymb]{revtex4-2}

\usepackage{graphicx}
\usepackage{dcolumn}
\usepackage{bm}
\usepackage{epstopdf}
\usepackage{upgreek}
\usepackage{amsmath,amssymb,amsthm,commath,esint,tikz-cd,tikz, mathtools}
\usepackage{array}

\numberwithin{definition}{section}

\numberwithin{proposition}{section}

\numberwithin{theorem}{section}

\numberwithin{lemma}{section}

\numberwithin{corollary}{section}
\theoremstyle{remark}

\numberwithin{example}{section}

\newcommand{\ad}{\textrm{ad}}

\usepackage{comment}

\begin{document}

\title{Ginzburg-Landau surface energy of multiband superconductors: Derivation and application to selected systems}

\author{J. Bekaert}
\email{jonas.bekaert@uantwerpen.be}
\affiliation{%
 Department of Physics \& NANOlab Center of Excellence, University of Antwerp,
 Groenenborgerlaan 171, B-2020 Antwerp, Belgium
}
\author{L. Bringmans}
\affiliation{%
 Department of Physics \& NANOlab Center of Excellence, University of Antwerp,
 Groenenborgerlaan 171, B-2020 Antwerp, Belgium
}
\author{M. V. Milo\v{s}evi\'{c}}
\affiliation{%
 Department of Physics \& NANOlab Center of Excellence, University of Antwerp,
 Groenenborgerlaan 171, B-2020 Antwerp, Belgium
}


\begin{abstract}
\noindent We determine the energy of an interface between a multiband superconducting and a normal half-space, in presence of an applied magnetic field, based on a multiband Ginzburg-Landau (GL) approach. We obtain that the multiband surface energy is fully determined by the critical temperature, electronic densities of states, and superconducting gap functions associated with the different band condensates. This furthermore yields an expression for the thermodynamic critical magnetic field, in presence of an arbitrary number of contributing bands. Subsequently, we investigate the sign of the surface energy as a function of material parameters, through numerical solution of the GL equations. Here, we consider two distinct cases: (i) standard multiband superconductors with attractive interactions, and (ii) a three-band superconductor with a chiral ground state with phase frustration, arising from repulsive interband interactions. Furthermore, we apply this approach to several prime examples of multiband superconductors, such as metallic hydrogen and MgB$_2$, based on microscopic parameters obtained from first-principles calculations.  

\end{abstract}

\maketitle

\section{Introduction}

The behavior of superconductors in an applied magnetic field has been an active area of research ever since Landau postulated the possibility of an intermediate state in which normal and superconducting regions coexist \cite{Landau1943}, followed by a generalized analysis of superconductor-normal interfaces by Ginzburg and Landau, applying the eponymous Ginzburg-Landau (GL) theory for superconductors \cite{GinzburgLandau1950}. To this end, they considered the energy associated with such interface -- called surface energy $\sigma_{\mathrm{ns}}$ -- demonstrating it to be fully parameterized by a single dimensionless parameter, the GL parameter $\kappa$ (as reviewed in textbooks, e.g., Refs. \citenum{Fetter_Walecka,Abrikosov_textbook}). Subsequently, Abrikosov explored the case where $\kappa>1/\sqrt{2}$, finding $\sigma_{\mathrm{ns}}<0$, hence, the system minimizes its total energy by creating many superconductor-normal interfaces, which is known as type-II superconductivity. The smallest possible normal regions are vortices, carrying a single magnetic flux quantum each, which can arrange in a vortex lattice \cite{Abrikosov1952,Abrikosov1957}. On the other hand, $\kappa<1/\sqrt{2}$ leads to $\sigma_{\mathrm{ns}}>0$, hence the creation of superconductor-normal interfaces is not energetically preferential in this case. Thus, a type-I phase transition from the superconducting to the normal state occurs at the thermodynamic critical field $H_{\mathrm{c}}$. This dichotomy between type-I and type-II superconductors, and its relation to GL parameter $\kappa$, has since played a central role in superconductivity research.

The extension of GL theory to systems with two dissimilar electronic bands at the Fermi level -- as is the case in e.g. certain transition metals -- was soon after explored \cite{Tilley_1964}. Subsequently, the discovery in 2001 of distinctly two-gap superconductivity in magnesium diboride (MgB$_2$) accompanied by an elevated critical temperature ($T_{\mathrm{c}}$) of 39 K \cite{Choi2002,Souma2003,Nagamatsu2001} -- featuring a stronger condensate stemming from its $\sigma$ bands and a weaker one from the $\pi$ bands -- sparked renewed, widespread interest in multiband and multigap superconductors \cite{Tanaka_2015}. Such two-gap superconductors have been proposed to enable opposite tendencies for short-range and long-range vortex-vortex interactions, potentially resulting in the formation of stripes and clusters of vortices \cite{PhysRevB.83.214523,PhysRevLett.102.117001,Brandt2011}. 

Furthermore, in three-gap systems and beyond it is not \textit{a priori} evident which phase minimizes the energy functional, especially in the presence of repulsive interband interactions. Attractive interactions favor the same phase between the condensates as their ground state, while repulsive interactions favor a phase difference of $\pi$ \cite{TANAKA20101980}. In case the repulsive interband interactions prevail over the intraband ones, one can thus obtain two superconducting gaps with opposite sign within a two-band model \cite{TANAKA20101980,PhysRevB.81.134522}. Such spin-singlet sign-changing $s$-wave gap symmetry in a multiband system is denoted as $s^{\pm}$ pairing. There is growing evidence that this is the pairing symmetry of iron-based superconductors like the pnictides \cite{Bang_2017}. 

In a three-band model with all-repulsive interband interactions one can furthermore obtain two degenerate, chiral solutions for the phases of the superconducting order parameters, accompanied by time-reversal symmetry breaking (TRSB) \cite{TANAKA20101980,PhysRevB.81.134522}. This can lead to spontaneous currents and fields wherever translational symmetry is broken in the sample (at edges, impurities, domain walls, ...). 

Since the discovery of MgB$_2$ as the first distinct two-gap superconductor, many systems hosting distinct multiband and multigap superconducting properties have been identified. In order to describe these new systems, GL theory has recently been extended to systems with an arbitrary number of bands, by means of a systematic Gor'kov truncation procedure applied to the multiband BCS Hamiltonian \cite{PhysRevB.87.134510}. Here, we apply this multiband GL functional to investigate the surface energy of a superconducting-normal interface, in the presence of an arbitrary number of superconducting band condensates.

The paper is organized as follows. In Sec.~\ref{sec:derivation}, we derive an analytical expression for the surface energy. Subsequently, in Sec.~\ref{sec:crit_field}, we obtain an expression for the corresponding thermodynamic critical magnetic field, which we also apply this to the case of metallic hydrogen, hosting three superconducting gaps. In Sec.~\ref{sec:non-degen}, we consider an $N$-band superconductors with purely attractive interactions between the bands, and characterize the sign of the surface energy depending on the materials parameters. We also demonstrate the application of this approach to MgB$_2$ in bulk and monolayer form, based on microscopic parameters obtained from first-principles calculations. Finally, in Sec.~\ref{sec:chiral}, we investigate the case of a chiral three-band superconductor with phase frustration, resulting from all-repulsive interband interactions. 

\section{Deriving a general expression for the multiband surface energy}
\label{sec:derivation}

To derive the analytical expression for $\sigma_{\mathrm{ns}}$, we generalize the textbook approach for the single-band case, e.g.~presented in Ref.~\citenum{Fetter_Walecka}. Here, we make use of the free energy density for multigap superconductors established in Ref.~\citenum{PhysRevB.87.134510},
\begin{align}
\mathcal{F}= \mathcal{F}_{\mathrm{n}0}+\sum_{\alpha \beta} a_{\alpha \beta} \psi_{\alpha}^{*}\psi_{\beta} + K_{\alpha \beta} \boldsymbol{\mathrm{D}}^{*}\psi_{\alpha}^{*} \boldsymbol{\mathrm{D}} \psi_{\beta} 
 + \frac{1}{2}\sum_{\alpha \beta \gamma \delta} b_{\alpha \beta \gamma \delta} \psi_{\alpha}^{*}\psi_{\beta} \psi_{\gamma}^{*}\psi_{\delta} +\frac{\boldsymbol{\mathrm{B}}^2}{8\pi}~,
\label{eq:func}
\end{align}
where $\mathcal{F}_{\mathrm{n}0}$ is the free energy in the normal state in the absence of an applied magnetic field, $\boldsymbol{\mathrm{D}}=\boldsymbol{\nabla}+i\frac{e^*}{\hbar c}\boldsymbol{\mathrm{A}}$ ($\boldsymbol{\mathrm{A}}$ being the magnetic vector potential) and $e^*=-2e$ is the Cooper pair charge. The indices $\alpha, \beta, \gamma, \delta$ run over $1,...,M$, where $M$ signifies the number of degenerate solutions of the gap equation to lowest order, $\check{L}\boldsymbol{\Delta}^{(0)}=0$, where $\boldsymbol{\Delta}^{(0)}\propto \sqrt{\tau}=\sqrt{1-\frac{T}{T_{\mathrm{c}}}}$, that yield the same maximal critical temperature $T_{\mathrm{c}}$.
Furthermore, the coefficients in the functional can be related to the following microscopic parameters: (i) the density of states \textit{per band} at the Fermi level $N_{\mathrm{F},i}$ and (ii) the average Fermi velocity per band $v_{\mathrm{F},i}$, by the following sums over band index $i=1,...,N$ (in CGS units), 
\begin{align}
\begin{cases}
a_{\alpha \beta}&=\sum\limits_i a_i \xi_{\alpha i} \xi_{\beta i} ~\mathrm{with}~ a_i=-N_{\mathrm{F},i} \tau~,\\
b_{\alpha \beta \gamma \delta}&=\sum\limits_i b_i \xi_{\alpha i} \xi_{\beta i} \xi_{\gamma i} \xi_{\delta i} ~\mathrm{with}~ b_i=N_{\mathrm{F},i}\cdot \frac{7 \zeta(3)}{8\pi^2T_{\mathrm{c}}^2}~,\\
K_{\alpha \beta}&=\sum\limits_i K_i \xi_{\alpha i} \xi_{\beta i} ~\mathrm{with}~ K_i=\frac{b_i}{6}\hbar^2 v_{\mathrm{F},i}^2~.
\end{cases}
\label{eq:coefficients}
\end{align}
The surface energy $\sigma_{\mathrm{ns}}$ at a flat 2D interface between normal material (at $z<0$) and the superconductor (at $z>0$) in an applied magnetic field parallel to the interface $\textbf{B}_{\mathrm{c}}=B_{\mathrm{c}} \hat{x}$, is defined as \cite{Fetter_Walecka}:
\begin{align}
\sigma_{\mathrm{ns}}=\int_{-\infty}^{+\infty} \left[ \mathcal{F}(z)-\frac{B(z)B_{\mathrm{c}}}{4\pi}-\mathcal{F}_{\mathrm{n}0} +\frac{\textbf{B}_{\mathrm{c}}^2}{8\pi}\right] dz~,
\label{eq:surfen}
\end{align}
where the sum of the first two terms is the Gibbs free energy for a superconductor. Filling in Eq.~\eqref{eq:func}, we obtain in the multiband case
\begin{align}
\sigma_{\mathrm{ns}}=\int_{-\infty}^{+\infty} \left[ \sum_{\alpha \beta} a_{\alpha \beta} \psi_{\alpha}^{*}\psi_{\beta} + K_{\alpha \beta} \boldsymbol{\mathrm{D}}^{*}\psi_{\alpha}^{*} \boldsymbol{\mathrm{D}} \psi_{\beta} + \frac{1}{2}\sum_{\alpha \beta \gamma \delta} b_{\alpha \beta \gamma \delta} \psi_{\alpha}^{*}\psi_{\beta} \psi_{\gamma}^{*}\psi_{\delta} +\frac{1}{8\pi}(\textbf{B}_{\mathrm{c}}-\textbf{B})^2\right] dz ~,
\end{align}
where we note that the last term depends on the difference between applied magnetic field $\textbf{B}_{\mathrm{c}}$ and magnetic field \textbf{B} in the superconductor. To proceed, we need to combine this with the first GL equation, being a set of $M$ equations in the multiband case,
\begin{align}
\sum_{\beta}\left(a_{\alpha \beta}-K_{\alpha \beta} \textbf{D}^2\right)\psi_{\beta}+\sum_{\beta \gamma \delta} b_{\alpha \beta \gamma \delta} \psi_{\beta} \psi_{\gamma}^*\psi_{\delta}=0~.
\end{align}
In order to use this expression, we can simplify it by applying the Coulomb gauge $\boldsymbol{\nabla}\cdot \textbf{A}=0$ to
\begin{align}
\textbf{D}^2\psi_{\beta}=\left(\boldsymbol{\nabla}^2+\frac{ie^*}{\hbar c}(\boldsymbol{\nabla}\cdot \textbf{A}+\textbf{A}\cdot \boldsymbol{\nabla})-\left(\frac{e^*}{\hbar c}\right)^2 \textbf{A}^2\right)\psi_{\beta}~.
\end{align}
Moreover, with magnetic field $\textbf{B}(z)=B(z) \hat{x}$, the vector potential is of the form $\textbf{A}(z)=A(z)\hat{y}$. In Ref.~\citenum{Fetter_Walecka}, it is demonstrated from the symmetry of the supercurrent $\textbf{j}(z)=\vert j(z) \vert \hat{y}$ that the order parameters can be written as $\psi_{\beta}=\mathrm{e}^{i\phi_{\beta}(y)} \vert\psi_{\beta}(z)\vert$, so in the case we consider here with full $xy$-symmetry of the condensates ($\phi_{\beta}(y)\equiv \phi_{\beta}$ a constant), $\textbf{A}\cdot \boldsymbol{\nabla}\psi_{\beta}=A(z)\frac{\partial}{\partial y}\psi_{\beta}=0$. As a result we obtain
\begin{align}
\textbf{D}^2\psi_{\beta}=\boldsymbol{\nabla}^2 \psi_{\beta}-\left(\frac{e^*}{\hbar c}\right)^2 \textbf{A}^2\psi_{\beta}~.
\end{align}
Thus we can restate the first GL equation as
\begin{align}
\sum_{\beta}\left(-K_{\alpha \beta}\psi_{\beta}''+K_{\alpha \beta}\left(\frac{e^*}{\hbar c}\right)^2\textbf{A}^2\psi_{\beta}+a_{\alpha \beta} \psi_{\beta}\right)+\sum_{\beta \gamma \delta} b_{\alpha \beta \gamma \delta}\psi_{\beta}\psi_{\gamma}^*\psi_{\delta}=0~,
\end{align}
where $\psi_{\beta}'=\partial \psi_{\beta}/\partial z$. Let us now integrate this expression with respect to $\psi_{\alpha}^* dz$. Using integration by parts, $\int_{-\infty}^{+\infty}\psi_{\alpha}^* \psi_{\beta}'' dz=\left[\psi_{\alpha}^*\psi_{\beta}'\right]_{-\infty}^{+\infty}-\int_{-\infty}^{+\infty}  \psi_{\alpha}^{*\prime} \psi_{\beta}' dz$, and since the condensates do not exist outside the superconductor and do not vary deep in the superconductor ($\psi_{\alpha}(-\infty)=0$ and $\psi_{\beta}'(+\infty)=0$), the first term vanishes. As a result, we end up with the following set of $M$ equations,
\begin{align}
\int_{-\infty}^{+\infty} \left[\sum_{\beta}\left(K_{\alpha \beta}\psi_{\alpha}^{* \prime}\psi_{\beta}'+K_{\alpha \beta}\left(\frac{e^*}{\hbar c}\right)^2\psi_{\alpha}^*\textbf{A}^2\psi_{\beta}+a_{\alpha \beta} \psi_{\alpha}^{*}\psi_{\beta}\right)+\sum_{\beta \gamma \delta} b_{\alpha \beta \gamma \delta}\psi_{\alpha}^{*}\psi_{\beta}\psi_{\gamma}^*\psi_{\delta}\right] dz=0~.
\label{eq:GL1}
\end{align}
We can now combine this with the previously obtained expression for the surface energy (Eq.~\eqref{eq:surfen}), using the Coulomb gauge, to obtain
\begin{align}
\sigma_{\mathrm{ns}}=\int_{-\infty}^{+\infty} \left[ \sum_{\alpha \beta} a_{\alpha \beta} \psi_{\alpha}^{*}\psi_{\beta} + K_{\alpha \beta}\psi_{\alpha}^{*\prime}\psi_{\beta}^{*\prime} \boldsymbol{\mathrm{D}}^{*}\psi_{\alpha}^{*} \boldsymbol{\mathrm{D}} \psi_{\beta} + \frac{1}{2}\sum_{\alpha \beta \gamma \delta} b_{\alpha \beta \gamma \delta} \psi_{\alpha}^{*}\psi_{\beta} \psi_{\gamma}^{*}\psi_{\delta} +\frac{1}{8\pi}(\textbf{B}_{\mathrm{c}}-\textbf{B})^2\right] dz ~.
\end{align}
Summing up Eqs.~\eqref{eq:GL1} over $\alpha=1,...,M$, we obtain
\begin{align}
\sigma_{\mathrm{ns}}=\int_{-\infty}^{+\infty} \left[-\frac{1}{2}\sum_{\alpha \beta \gamma \delta} b_{\alpha \beta \gamma \delta} \psi_{\alpha}^{*}\psi_{\beta} \psi_{\gamma}^{*}\psi_{\delta} +\frac{1}{8\pi}(\textbf{B}_{\mathrm{c}}-\textbf{B})^2\right] dz~,
\label{eq:surfE}
\end{align}
To relate this expression to experimental results, it is advantageous to rewrite it in terms of the superconducting gap functions $\Delta_i$. To this end, we introduce $b_{\alpha \beta \gamma \delta}=\displaystyle\sum_{i=1}^N b_i \xi_{\alpha i} \xi_{\beta i} \xi_{\gamma i} \xi_{\delta i}=\sum_{i=1}^N \frac{7 \zeta(3)}{8\pi^2 T_{\mathrm{c}}^2}N_{\mathrm{F},i} \xi_{\alpha i} \xi_{\beta i} \xi_{\gamma i} \xi_{\delta i}$. Therefore, we obtain for the first term of the integrand of $\sigma_{\mathrm{ns}}$,
\begin{align}
-\frac{1}{2}\sum_{\alpha \beta \gamma \delta} b_{\alpha \beta \gamma \delta} \psi_{\alpha}^{*}\psi_{\beta} \psi_{\gamma}^{*}\psi_{\delta}=\frac{-7 \zeta(3)}{16\pi^2 T_{\mathrm{c}}^2}\sum_{i=1}^N N_{\mathrm{F},i}\left(\sum_{\alpha=1}^M \psi_{\alpha}^* \xi_{\alpha i}\right) \left(\sum_{\beta=1}^M \psi_{\beta}\xi_{\beta i}\right) \left(\sum_{\gamma=1}^M \psi_{\gamma}^* \xi_{\gamma i}\right) \left(\sum_{\delta=1}^M \psi_{\delta} \xi_{\delta i}\right)~.
\label{eq:term}
\end{align}
The properties of $\xi_{\alpha i}$ are dictated by the matrix $L_{ij}$ of which $\xi_{\alpha i}$ are the eigenvectors. $L_{ij}=\delta_{ij}\left(\gamma_{ii}-N_{F,i}\mathcal{A}\right)+(1-\delta_{ij})\gamma_{ij}$ clearly contains only real elements, moreover it is a symmetric matrix, i.e. $L_{ij}=L_{ji}$. This follows from the form of its off-diagonal elements $(1-\delta_{ij})\gamma_{ij}$. $\gamma_{ij}$ are the elements of the inverse of the coupling matrix, which is symmetric. The inverse of a symmetric matrix is moreover also symmetric. The eigenvectors of the real and symmetric matrix $L_{ij}$ can always be chosen to be real. So, it follows from linear algebra that $\xi_{\alpha i}$ are always real. This enables rewriting Eq.~\eqref{eq:term} using $\boldsymbol{\Delta}^{(0)}=\sum_{\alpha=1}^M \psi_{\alpha} \boldsymbol{\xi}_{\alpha}$ (the lowest order, $\propto \sqrt{\tau}$, in the expansion $\boldsymbol{\Delta}=\boldsymbol{\Delta}^{(0)}+\boldsymbol{\Delta}^{(1)}+\mathcal{O}
(\tau^{5/2})$), yielding $\frac{-7 \zeta(3)}{16\pi^2 T_{\mathrm{c}}^2}\sum_{i=1}^N N_{\mathrm{F},i} \vert \Delta_i^{(0)} \vert^4$. Therefore, we obtain as final result,
\begin{align}
\sigma_{\mathrm{ns}}=\int_{-\infty}^{+\infty} \left[\frac{-7 \zeta(3)}{16\pi^2 T_{\mathrm{c}}^2}\sum_{i=1}^N N_{\mathrm{F},i} \vert \Delta_i^{(0)}(z) \vert^4 +\frac{1}{8\pi}(\textbf{B}_{\mathrm{c}}-\textbf{B}(z))^2\right] dz ~.
\label{eq:result}
\end{align}

\section{Multiband thermodynamic critical magnetic field}
\label{sec:crit_field}

The thermodynamic critical magnetic field of multiband superconductors immediately follows from Eq.~\eqref{eq:result}, through the condition that \textbf{B} has to vanish deep within the superconducting region (i.e., $z\rightarrow \infty$). This yields
\begin{align}
B_{\mathrm{c}}=\sqrt{\frac{7\zeta(3)}{2\pi T_{\mathrm{c}}^2}\sum_{i=1}^NN_{\mathrm{F},i}\vert \Delta_{i,\infty}^{(0)}\vert^4}~,
\label{eq:crit_field}
\end{align}
up to order $\tau$. This expression provides a separate route to calculate the critical magnetic field compared with other approaches, for which the complete band-resolved electron-phonon coupling matrix has to be known \cite{PhysRevB.85.014502}. 

\subsection{Application to metallic hydrogen}
\label{subsec:metallicH}

As a direct application, we evaluate the critical magnetic field of metallic hydrogen under ultrahigh pressure, first predicted to be a high-$T_{\mathrm{c}}$ superconductor by Ashcroft \cite{PhysRevLett.21.1748}. More detailed \textit{ab initio} calculations within the density functional theory for the superconducting state (SCDFT) framework have identified different structural phases of solid hydrogen, stabilized in different pressure regimes \cite{PhysRevB.81.134505}. 

The metallic \textit{Cmca}-phase, with a a base-centered orthorhombic (bco) unit cell containing two H$_2$ molecules, is stable in the pressure range from 400 to 500 GPa \cite{PhysRevLett.100.257001,PhysRevB.81.134505}. The calculated $T_{\mathrm{c}}$ values range from 84 K at 414 GPa to 242 K at 450 GPa \cite{PhysRevLett.100.257001,PhysRevB.81.134506}. At 414 GPa, SCDFT calculations have revealed the presence of three groups of Fermi pockets with different superconducting gap values \cite{PhysRevLett.100.257001}. $\Delta_1$ is the strongest gap (average of 19.3 meV at $T=0$). $\Delta_2$ and $\Delta_3$ overlap between 13 meV and 15.8 meV at $T=0$, but their average values are nevertheless distinct (15.4 meV and 13.6 meV respectively). Therefore, this system can be described as a three-gap superconductor within our GL description.   

\begin{figure}[t]
\centering
\includegraphics[width=0.5\linewidth]{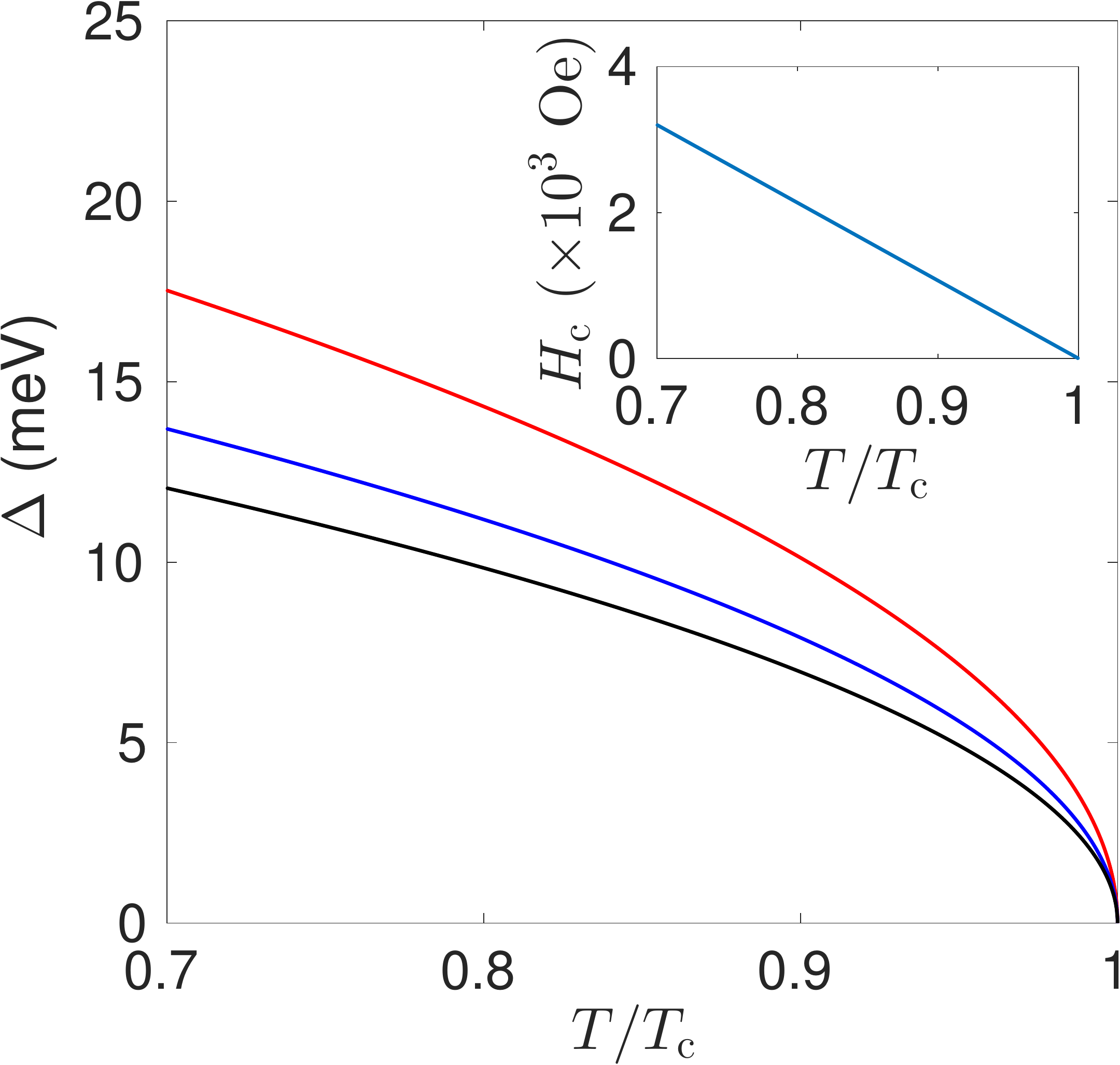}
\caption{(Color online) GL evolution of the averages of the three gaps of metallic hydrogen with temperature, fitted from Ref.~\citenum{PhysRevLett.100.257001}. The inset shows the resulting temperature-evolution of the thermodynamic critical magnetic field $H_{\mathrm{c}}$ calculated using Eq.~\eqref{eq:crit_field}.}
\label{fig:crit_field_hydrogen}
\end{figure}

The temperature-evolution of the three superconducting gap values within the GL description, fitted from the SCDFT result in the temperature range $\left[0.7 - 1 \right] T_{\mathrm{c}}$, is shown in Fig.~\ref{fig:crit_field_hydrogen}. Based on these gap values and the partial DOS values of the different bands (see Table \ref{tab:hydrogen}), we evaluate the thermodynamic critical magnetic field, and its temperature dependence in the vicinity of $T_{\mathrm{c}}$, of the \textit{Cmca}-phase at 414 GPa using the multiband expression derived above (Eq.~\eqref{eq:crit_field}). The result is depicted in the inset of Fig.~\ref{fig:crit_field_hydrogen}. Our calculations show that $H_{\mathrm{c}}$ attains elevated values, reaching 3194 Oe at $0.7 T_{\mathrm{c}}$, owing to the large superconducting gap values in this system. 

\section{Type-I/Type-II behavior in non-degenerate $N$-band superconductors}
\label{sec:non-degen}

\subsection{Determining the multiband Ginzburg-Landau parameter}
\label{subsec:non-degen}

\begin{figure}[t]
    \includegraphics[width = 0.6\linewidth]{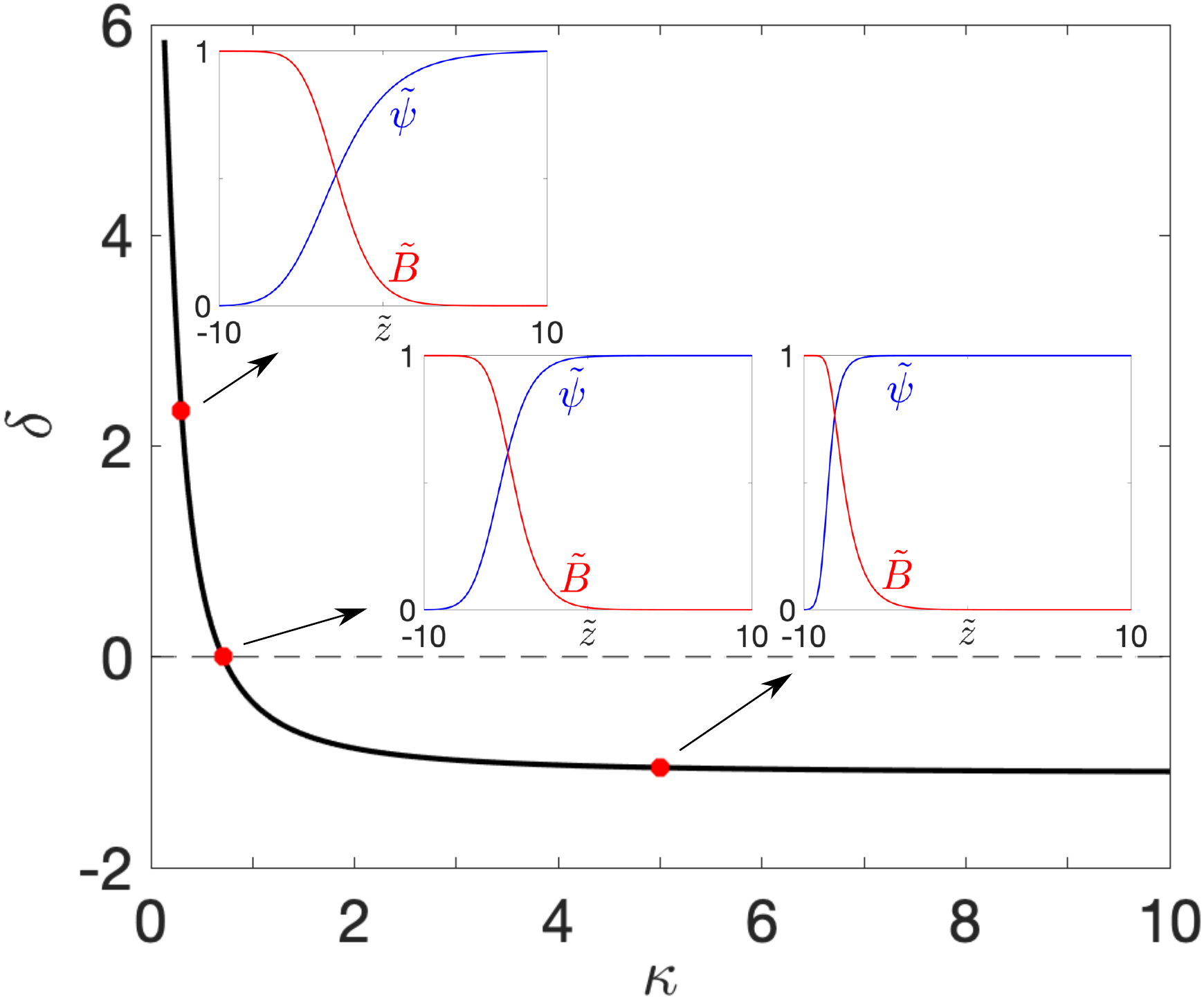}
    \caption{(Color online) Dimensionless surface energy parameter $\delta$ ($=8 \pi \sigma_{\textrm{ns}}/(\delta_0B_{\mathrm{c}}^2)$) as a function of the Ginzburg-Landau parameter $\kappa$ (black line). The insets show the solutions $\tilde{\psi}$ and $\tilde{B}$ of the dimensionless Ginzburg-Landau equations for $\kappa = 0.3,~ 1/\sqrt{2},~ 5$ (red dots). The dashed line $(\delta=\sigma_{\textrm{ns}}=0)$ intersects the curve $\delta(\kappa)$ at $\kappa=1/\sqrt{2}$.}
    \label{fig_surf_en_multiband}
 \end{figure}

The aim is to first rewrite the GL equations for non-degenerate $N$-band superconductors (the case where $M=1$) in a dimensionless form, in the setting of a normal-superconducting half space:
\begin{align}
\begin{cases}
-K\frac{d^2\psi}{dz^2}+\displaystyle\left(a+K\left(\frac{e^*}{\hbar c}\right)^2A^2\right)\psi+b\abs{\psi}^2\psi=0~,\\
\frac{d^2A}{dz^2} = \vspace{0.3cm}\displaystyle 8\pi\left(\frac{e^*}{\hbar c}\right)^2K\abs{\psi}^2A~,
\end{cases}
\label{unscaled}
\end{align}
accompanied by the boundary conditions
\begin{align}
    \begin{cases}
    \psi =0, ~B=B_{\mathrm{c}}, \qquad z\to -\infty~,\\
    \psi =\abs{\psi_\infty}, ~B=0, \qquad z\to +\infty~.
    \end{cases}
\end{align}
In accordance with Ref.~\citenum{GinzburgLandau1950}, we define the following dimensionless quantities:
\begin{align}
\tilde{\psi}=\frac{\psi}{\abs{\psi_\infty}}~, ~ \tilde{A}=\sqrt{\left(\frac{e^*}{\hbar c}\right)^2\frac{K}{-a}}A~,~
\delta_0^2= \frac{1}{8\pi K\abs{\psi_\infty}^2}\left(\frac{\hbar c}{e^*}\right)^2~,~ \kappa^2= \frac{b}{8\pi K^2}\left(\frac{\hbar c}{e^*}\right)^2~,~\tilde{z}= \frac{z}{\delta_0}~.
\label{def_dimensionless}
\end{align}
Direct application of the chain rule gives the GL equations in dimensionless form, 
\begin{align}
\begin{cases}
\frac{d^2\tilde{\psi}}{d\tilde{z}^2}&= \kappa^2\left((-1+\tilde{A}^2)\tilde{\psi}+\lvert\tilde{\psi}\rvert^2\tilde{\psi}\right)~,
\\
\frac{d^2\tilde{A}}{d\tilde{z}^2} &=\lvert\tilde{\psi}
\rvert^2\tilde{A}~.
\end{cases}
\label{eq:multibandGL}
\end{align}
To determine the corresponding boundary conditions we need to calculate $\tilde{B}$, the magnetic field corresponding to the magnetic vector potential $\tilde{A}$. First, we compute the critical magnetic field $B_{\mathrm{c}}$. For notational convenience we set $ C= 7\zeta(3)/(16\pi^2 T_{\mathrm{c}}^2)$. Since
\begin{align}b=2C\sum_{i=1}^N N_{\mathrm{F},i}\abs{\xi_i}^4~,
\end{align}
we can rewrite the critical magnetic field $B_{\mathrm{c}}$ as
\begin{align}B_{\mathrm{c}}^2 =8\pi C\sum_{i=1}^NN_{\mathrm{F},i}|\psi_\infty|^4\abs{\xi_i}^4=4\pi b\abs{\psi_\infty}^4=\frac{4\pi a^2}{b}~.
\end{align}
The computation of $\tilde{B}$ from Eq.~\eqref{def_dimensionless} is then straightforward:
\begin{align}\tilde{B}=\frac{d\tilde{A}}{d\tilde{z}}=\frac{B}{\sqrt{2}B_{\mathrm{c}}}~.
\end{align}
The accompanying dimensionless boundary conditions for Eq.~\eqref{eq:multibandGL} read
\begin{align}\label{eq:multibandBC}
\begin{cases}
    \tilde{\psi} =0,~ \tilde{B}=\displaystyle\frac{1}{\sqrt{2}}, \qquad \tilde{z}\to -\infty,\\
    \tilde{\psi} =1,~ \tilde{B}=0, \qquad \tilde{z}\to +\infty.
    \end{cases}
\end{align}
This form is convenient for direct numerical simulations of the GL equations. Evaluation of the surface energy requires a reparametrization of the integral in Eq.~\eqref{eq:surfE} as a function of the rescaled order parameter $\tilde{\psi}$ and magnetic field $\tilde{B}$.
A straightforward computation gives
\begin{align}\sigma_{\textrm{ns}}=\frac{b}{2}\delta_0\abs{\psi_\infty}^4\int_{-\infty}^{+\infty} \left[-\vert\tilde{\psi}(\delta_0\tilde{z})\vert^4+(\sqrt{2}\tilde{B}(\delta_0\tilde{z})-1)^2 \right] d\tilde{z}= \frac{\delta_0B_{\mathrm{c}}^2}{8\pi}\delta~,
\end{align}
where $\delta$ is the dimensionless quantity defined by
\begin{align}\label{eq:SEmulti}
\delta= \int_{-\infty}^{+\infty} \left[-\vert\tilde{\psi}(\delta_0\tilde{z})\vert^4+(\sqrt{2}\tilde{B}(\delta_0\tilde{z})-1)^2 \right] d\tilde{z}~.
\end{align}
Once $\tilde{\psi}$ and $\tilde{B}$ are obtained from the GL Eq.~\eqref{eq:multibandGL} for any value of $\kappa$, this integral can be numerically computed. The numerical procedures we have used to do this are described in Appendix \ref{appendix_num_meth}. The result for $\delta$ as a function of $\kappa$ is shown in Fig.~\ref{fig_surf_en_multiband}. The insets show solutions for $\tilde{\psi}$ and $\tilde{B}$ for selected $\kappa$ values. As for the single-band case, the solution has two regimes. For $\kappa<1/\sqrt{2}$, $\delta$ ($=8 \pi \sigma_{\textrm{ns}}/(\delta_0B_{\mathrm{c}}^2)$) is positive (type-I regime), while for $\kappa>1/\sqrt{2}$, $\delta$ is negative (type-II regime). Hence, the non-degenerate multiband case (with $M=1$) can be parameterized using a single GL parameter $\kappa$, regardless of the number of bands $N$, and the resulting behavior of the surface energy maps exactly to the single-band case within the Ginzburg-Landau regime described by the functional in Eq.~\eqref{eq:func}. This generalizes the conclusion obtained for two-band superconductors by Geyer \textit{et al.}, albeit based on a different GL functional with Josephson-type coupling between the two bands \cite{PhysRevB.82.104521}. 

\subsection{Application to bulk and monolayer magnesium diboride}
\label{subsec:application_MgB2}

As mentioned in the Introduction, magnesium diboride (MgB$_2$) is a prototype multigap superconductor. Bulk MgB$_2$ hosts two distinct superconducting gaps: the $\sigma$ gap and the $\pi$ gap, stemming from $\sigma$ bonds of boron-$p_{x,y}$ orbitals and $\pi$ bonds of boron-$p_z$ orbitals respectively \cite{Choi2002}. MgB$_2$ has been predicted to develop another distinct band composed mainly of Mg-$p$ orbitals in the atomically-thin limit \cite{PhysRevB.96.094510,Bekaert2017}. This state is localized at the free magnesium surface of the multilayer structure, hence it was named `surface state' ($S$) \cite{PhysRevB.96.094510,Bekaert2017}. In the case of a single monolayer (ML) of MgB$_2$, the three gaps -- $\sigma$, $\pi$ and $S$ -- are fully separated, giving rise to distinct three-gap superconductivity \cite{PhysRevB.96.094510}. The critical temperatures are $T_{\mathrm{c}} = 39$ K for bulk MgB$_2$ \cite{Choi2002,Souma2003,Nagamatsu2001}, and 20 K for ML MgB$_2$, the latter obtained from \textit{ab initio} calculations within the Eliashberg framework \cite{PhysRevB.96.094510}. 

In both the bulk and ML case the gaps are non-degenerate, hence $M=1$. In Section \ref{subsec:non-degen} we have demonstrated that this case is described by a single GL parameter $\kappa$, which depends on microscopic parameters like Fermi velocities, electronic DOS at $E_{\mathrm{F}}$ and the electron-phonon coupling matrix. The full set of microscopic parameters for bulk and ML MgB$_2$ is provided in Appendix \ref{AppendixC}. These were obtained from our prior density functional theory (DFT) and density functional perturbation theory (DFPT) results -- computational details are provided in Refs.~\citenum{PhysRevB.94.144506} and \citenum{PhysRevB.96.094510}, for bulk and ML MgB$_2$ respectively. The superconducting length scales at $T=0$ and GL parameter $\kappa$ are determined from these microscopic values using \cite{PhysRevB.87.134510}, cf.~Eq.~\eqref{def_dimensionless},
\begin{align}
\lambda_{\mathrm{L}}(0)=\frac{\hbar c}{\abs{e^*}}\sqrt{\frac{b}{8 \pi K \abs{a}}}~,~\xi(0)=\sqrt{\frac{K}{\abs{a}}}~,~\kappa=\frac{\lambda_{\mathrm{L}}}{\xi}=\frac{\hbar c}{\abs{e^*}}\sqrt{\frac{b}{8 \pi K^2}}~.
\end{align}
Here, $a$, $b$ and $K$ are calculated from the expressions in Eq.~\eqref{eq:coefficients} using the eigenvector $\boldsymbol{\xi}$ of $\check{L}$, having 2 components for bulk MgB$_2$ and 3 components for ML MgB$_2$. The results are summarized in Table \ref{Tab_results_MgB2}. 

Calculation of the superconducting length scales of bulk MgB$_2$ for each band condensate separately, using the first-principles values stated in Table \ref{tab:bulk_mgb2}, yields $\kappa_{\sigma}=2.61$ and $\kappa_{\pi}=0.71$. This corroborates the large discrepancy in nominal length scales of the two band condensates in bulk MgB$_2$ reported earlier \cite{PhysRevLett.102.117001}. We obtain $\kappa=1.64$ as the overall GL parameter of bulk MgB$_2$. Therefore, $\sigma_{\mathrm{ns}} < 0$, showing type-II behavior from the merger of both band condensates.

Analogously, for ML MgB$_2$ we obtain $\kappa=0.65$, so $\sigma_{\mathrm{ns}} > 0$. This marked reduction of $\kappa$ towards the ML limit is dominated by the increase of the coherence length according to $\xi(0) \propto T_{\mathrm{c}}^{-1}$, as $T_{\mathrm{c}}$ of the ML case is nearly a factor of 2 lower than that of the bulk. In addition, the increase of the average Fermi velocity and partial DOS of the leading $\sigma$ component (see Appendix \ref{AppendixC}) further reduce $\kappa$.  

We note that we have focused here on how intrinsic differences in the microscopic parameters between bulk and ML MgB$_2$ affect the multiband GL parameter. A more detailed analysis of the ML case would entail the dependence of $\psi$ and $\boldsymbol{B}$ on both $z$ and the out-of-plane direction, which goes beyond the 1D description (as a function of the $z$ coordinate) developed in this work. 

\begin{table}[t]
\centering
\begin{tabular}{| c| c| c | c | c |} 
\hline
Compound & $\lambda_{\mathrm{L}}(0)$ (nm) & $\xi(0)$ (nm) & $\kappa$  \\
 \hline
Bulk MgB$_2$ & $26.6$ & $16.2$ & 1.64\\
\hline
ML MgB$_2$ & $19.2$ & $29.4$  & 0.65 \\
\hline
\end{tabular}
\caption{Calculated superconducting length scales at $T=0$ and GL parameter $\kappa$ for bulk and monolayer MgB$_2$.}
\label{Tab_results_MgB2}
\end{table}

\section{Three-band chiral superconductor with phase frustration}
\label{sec:chiral}

Now we move to a particular three-band system with strong repulsive interband coupling, described by the coupling matrix $G_{ij}=g\left( 1-\delta_{ij}\right)$, where $g<0$ and $\delta_{ij}$ is the Kronecker delta \cite{PhysRevB.87.134510}. The inverse of this coupling matrix is $G^{-1}_{ij}=\gamma_{ij}=(-1)^{\delta_{ij}}/(2g)$. We will furthermore work within the assumption that all three bands have the same DOS at $E_{\mathrm{F}}$ ($N_{\mathrm{F}}$). The resulting gap equation $\check{L}\boldsymbol{\Delta}^{(0)} = \boldsymbol{0}$ only has non-trivial solutions provided that $\det\check{L}$ vanishes:
\begin{align}
    \det\check{L}  & = \frac{1}{2g}\det\begin{pmatrix} -1-2g N_{\mathrm{F}} \mathcal{A} & 1 & 1 \\ 1 & -1-2g N_{\mathrm{F}} \mathcal{A} & 1 \\ 1 & 1 & -1-2g N_{\mathrm{F}} \mathcal{A} 
\end{pmatrix}\\
& = -\frac{1}{2g}(2g N_{\mathrm{F}} \mathcal{A}+2)^2(2g N_{\mathrm{F}} \mathcal{A}-1)=0~,
\end{align}
where $\mathcal{A}=\mathrm{ln}\left( 2\mathrm{e}^{\Gamma} \hbar \omega_{\mathrm{c}}/(\pi T_{\mathrm{c}})\right)$ (with $\Gamma$ the Euler constant and $\omega_{\mathrm{c}}$ the characteristic cutoff frequency of the pairing) \cite{PhysRevB.87.134510}. The solutions are $\mathcal{A}_-=-1/(g N_{\mathrm{F}})$ and $\mathcal{A}_+=1/(2 g N_{\mathrm{F}})$, where the former has multiplicity 2. The smallest solution, $\mathcal{A}_{-}$, which yields the maximal critical temperature $T_{\mathrm{c}}= 2 \mathrm{e}^{\Gamma} \pi^{-1} \hbar \omega_{\mathrm{c}} \mathrm{exp}(-\mathcal{A}_-)$, is the solution that minimizes the energy functional. Hence, this system is characterized by degeneracy $M=2$. These two degenerate solutions are characterized by phase shifts of $\pm 2 \pi/3$ between the components of $\boldsymbol{\Delta}^{(0)}$ \cite{PhysRevB.87.134510}. They are chiral as they cannot be related by a rotation, and as a result foster time-reversal symmetry breaking (TRSB). 

By orthogonality of the vectors $\boldsymbol{\xi}_\alpha$ it follows that $K_{12}=K_{21}=a_{12}=a_{21}=0$. This allows us to define $\mathcal{K}_1= K_{11},K_2= K_{22}$ without ambiguity. Moreover, the equality of the DOS values implies that the tensor $b_{\alpha\beta\gamma\delta}$ is symmetric. Hence, we can reduce the notation to five independent values
\begin{align}
  \begin{split}
\beta_1:=b_{1111},~\beta_2=b_{1112}=b_{1121}=b_{1211}=b_{2111}~,\\
\beta_3:=b_{1122}=b_{1212}=b_{2112}=b_{2121}=b_{1221}=b_{2211}~,\\
\beta_4:=b_{1222}=b_{2122}=b_{2212}=b_{2221},~ \beta_5:=b_{2222}~.
  \end{split}
\end{align}
The two eigenvectors of $\check{L}$ corresponding to the eigenvalue $\mathcal{A}_-$ are
\begin{align}\boldsymbol{\xi}_1=\begin{pmatrix}0\\-1\\1\end{pmatrix},~\boldsymbol{\xi}_2=\begin{pmatrix}2\\-1\\-1\end{pmatrix}~.
\label{eq:chireig}
\end{align}
Using these expressions we can reduce the number of constants further to
\begin{align}\mathcal{K}_2=3\mathcal{K}_1,~ \alpha_2=3\alpha_1,~ \beta_2=\beta_4=0,~ \beta_3 = \beta_1,~ \beta_5=9\beta_1~.
\end{align}
Hence, we can rewrite the Ginzburg-Landau equations using only the constants $\alpha_1,\beta_1,\mathcal{K}_1$:
\begin{align}
\begin{cases}
\beta_1\left(\abs{\psi_1}^2+2\abs{\psi_2}^2+\psi_2^2\frac{\psi_1^*}{\psi_1}\right)+\alpha_1+\mathcal{K}_1\left(-\frac{\psi_1''}{\psi_1}+\left(\frac{e^*}{\hbar c}\right)^2A^2\right)=0~,\\
\beta_1\left(9\abs{\psi_2}^2+2\abs{\psi_1}^2+\psi_1^2\frac{\psi_2^*}{\psi_2}\right)+3\alpha_1+3\mathcal{K}_1\left(-\frac{\psi_2''}{\psi_2}+\left(\frac{e^*}{\hbar c}\right)^2A^2\right)=0~,\\
A''(z) = 8\pi \left(\frac{e^*}{\hbar c}\right)^2A(z)\mathcal{K}_1\left(|\psi_1(z)|^2+3|\psi_2(z)|^2\right)~.
\end{cases}
\label{eq:GL_chir_nonsymm}
\end{align}
Next, to fully exploit the symmetry of these GL equations, we introduce the following notation:
 \begin{align}\boldsymbol{\tilde{\psi}}=\begin{pmatrix}\tilde{\psi}_1\\  \tilde{\psi}_2\end{pmatrix}:=\frac{e^{-i\phi_1}}{\left|\psi_{\infty}\right|}\begin{pmatrix}\psi_1\\ \sqrt{3} \psi_2\end{pmatrix}~,~ \mathcal{D}:=\begin{pmatrix}
\tilde{\psi}_1 & 0 \\ 0 & \tilde{\psi_2}
\end{pmatrix}~,~ \mathcal{D}^{\textrm{ad}}:=\begin{pmatrix}
\tilde{\psi}_2 & 0 \\ 0 & \tilde{\psi}_1
\end{pmatrix}~,
\end{align}
where `ad' stands for the adjugate matrix. In this expression, rescaling $\psi_2$ by $\sqrt{3}$ symmetrizes the first two GL equations in Eq.~\eqref{eq:GL_chir_nonsymm} with respect to the permutation $\psi_1 \leftrightarrow \psi_2$. This facilitates the use of one vector equation for $\boldsymbol{\tilde{\psi}}$. Furthermore, the prefactor $e^{-i\phi_1}\left|\psi_{\infty}\right|^{-1}$ makes $\boldsymbol{\tilde{\psi}}$ dimensionless, and facilitates the treatment of the superconducting half-space and implementation of the boundary conditions (as elaborated below). Here, $\phi_1$ is the phase of $\psi_1$ deep in the superconducting region and $\left|\psi_{\infty}\right|^2=-3\alpha_1/(4\beta_1)=3|\alpha_1|/(4\beta_1)$
is the corresponding order parameter (identical for both components after rescaling the second component by $\sqrt{3}$). Note that the GL equations are invariant under the transformation $\boldsymbol{\psi}\mapsto e^{-i\phi_1}\boldsymbol{\psi}$. 

In analogy with the previous section, we define the following dimensionless quantities:
\begin{align}
\tilde{A}^2:=\frac{\mathcal{K}_1}{\abs{\alpha_1}}\left(\frac{e^*}{\hbar c}\right)^2A^2~,~\delta_0^2:=\frac{1}{8\pi\mathcal{K}_1\abs{\psi_\infty}^2}\left(\frac{\hbar c}{e^*}\right)^2~,~ \kappa^2:=\frac{\abs{\alpha_1}\delta_0^2}{\mathcal{K}_1}=\frac{\beta_1}{6\pi\mathcal{K}_1^2}\left(\frac{\hbar c}{e^*}\right)^2~,~ \tilde{z}:=\frac{z}{\delta_0}~.
\label{eq:dim_par_chiral}
\end{align}
The dimensionless constant $\kappa$ plays the same role here as the regular Ginzburg-Landau parameter in the non-degenerate multiband case treated in the previous section. The final symmetrized dimensionless form of the GL equations for the chiral case, which only includes $\kappa$ as a material-specific parameter, is
\begin{align}
\begin{cases}
\frac{d^2\tilde{\boldsymbol\psi}}{d\tilde{z}^2} &=\kappa^2\left[\frac{1}{4}\left(3\mathcal{D}\mathcal{D}^*+2\mathcal{D}^\ad(\mathcal{D}^\ad)^*\right)\tilde{\boldsymbol{\psi}}+\frac{1}{4}(\mathcal{D}^\ad)^2\tilde{\boldsymbol{\psi}}^*+(-1+ \tilde{A}^2)\tilde{\boldsymbol{\psi}}\right]~,\\
\frac{d^2\tilde{A}}{d\tilde{z}^2} &= \tilde{A}|\tilde{\boldsymbol{\psi}}|^2~.
\end{cases}
\label{eq:chirdimless}
\end{align}
To derive the accompanying boundary conditions, the magnetic field $\tilde{B}$ corresponding to the vector potential $\tilde{A}$ and critical magnetic field $B_{\mathrm{c}}$ need to be computed. First observe that the solution of the gap equation deep in the superconducting region is \cite{PhysRevB.87.134510}
\begin{align}\boldsymbol{\Delta}_\infty^{(0)}=\pm i\sqrt{\frac{\abs{\alpha_1}}{\beta_1}}\begin{pmatrix}1 \\ e^{\pm 2\pi i/3} \\ e^{-(\pm 1)2\pi i/3}\end{pmatrix}~,
\end{align}
with phase shifts between the components of $\pm 2 \pi/3$, as stated above. Both values yield the same $B_{\mathrm{c}}$ and surface energy, so they can be treated interchangeably here. The critical magnetic field is given by
\begin{align}B_{\mathrm{c}}^2=8\pi C N_{\mathrm{F}}\sum_{i=1}^3|\Delta_{i,\infty}^{(0)}|^4=\frac{6\pi\alpha_1^2}{\beta_1}~.
\end{align}
Using this value we can succinctly write $\tilde{B}$ as
\begin{align}\tilde{B}=\frac{d\tilde{A}}{d\tilde{z}}=\frac{B}{B_{\mathrm{c}}}~.
\end{align}
For numerical solution, explicit boundary conditions for the real and imaginary components of $\tilde{\psi}_1$ and $\tilde{\psi}_1$ need to be provided. Let $\phi_1, \phi_2$ be the phase arguments of $\tilde{\psi}_{1,\infty}$ and $\tilde{\psi}_{2,\infty}$ respectively. The phase difference $\delta \phi= \phi_2 - \phi_1$ has two solution branches: $\delta \phi = \pi/2+k2\pi$ and $\delta \phi = 3\pi/2+k2\pi$ with $k\in\mathbb{N}$ \cite{PhysRevB.87.134510}. Therefore, the appropriate boundary conditions are
\begin{align}\label{eq:chiralBC}
\begin{cases}
    \tilde{\boldsymbol{\psi}}=\mathbf{0}~,~ \tilde{B}=1 , \qquad \qquad ~~~ \tilde{z}\to -\infty,\\
    \tilde{\boldsymbol{\psi}}=\begin{pmatrix}1\\e^{i\delta\phi}\end{pmatrix}~,~ \tilde{B}=0,\qquad \tilde{z}\to+\infty~,
\end{cases}
\end{align}
where $e^{i\delta\phi}$ can only be $\pm i$, and $\tilde{z}\to -\infty$ corresponds to the region far outside of the superconductor and $\tilde{z}\to +\infty$ to the region deep inside the superconductor. The quantities needed to compute the surface energy are $|\tilde{\boldsymbol{\psi}}|^2$ and $\tilde{A}$ which are therefore identical for both choices of boundary conditions. Details on the implementation of these boundary conditions for complex $\tilde{\psi}_{1}$ and $\tilde{\psi}_{2}$ are provided in Appendix \ref{appendix_BCs_chiral}. 

To evaluate the surface energy according to the general formula in Eq.~\eqref{eq:result}, the eigenvectors provided in Eq.~\eqref{eq:chireig} allow explicit calculation of the components of $\boldsymbol{\Delta}^{(0)}$:
\begin{align}
\left|\Delta_{1}^{(0)}\right|^4=16|\psi_2|^4~,~\left|\Delta_{2}^{(0)}\right|^4=|\psi_1+\psi_2|^4~,~ \left|\Delta_{3}^{(0)}\right|^4=|\psi_1-\psi_2|^4~.
\end{align}
The surface energy of the chiral three-band case is therefore
\begin{align}
\sigma_{\textrm{ns}} 
=-C N_{\mathrm{F}} \int_{-\infty}^{+\infty}\left[\sum_{i=1}^3\vert \Delta_i^{(0)}(z) \vert^4+\frac{B_c^2}{8\pi}(\tilde{B}-1)^2\right]dz
= \frac{\delta_0\beta_1}{4}\abs{\psi_\infty}^4\delta~,
\end{align}
where
\begin{align} \label{eq:SEchir}
\delta=\int_{-\infty}^{+\infty}\left[-\frac{16}{9}|\tilde{\psi}_2|^4-\frac{1}{9}|\sqrt{3}\tilde{\psi}_1+\tilde{\psi}_2|^4-\frac{1}{9}|\sqrt{3}\tilde{\psi}_1-\tilde{\psi}_2|^4+4(\tilde{B}-1)^2\right] d\tilde{z}~.
\end{align}
\begin{figure}[t]
  \centering
    \includegraphics[width = 0.6\linewidth]{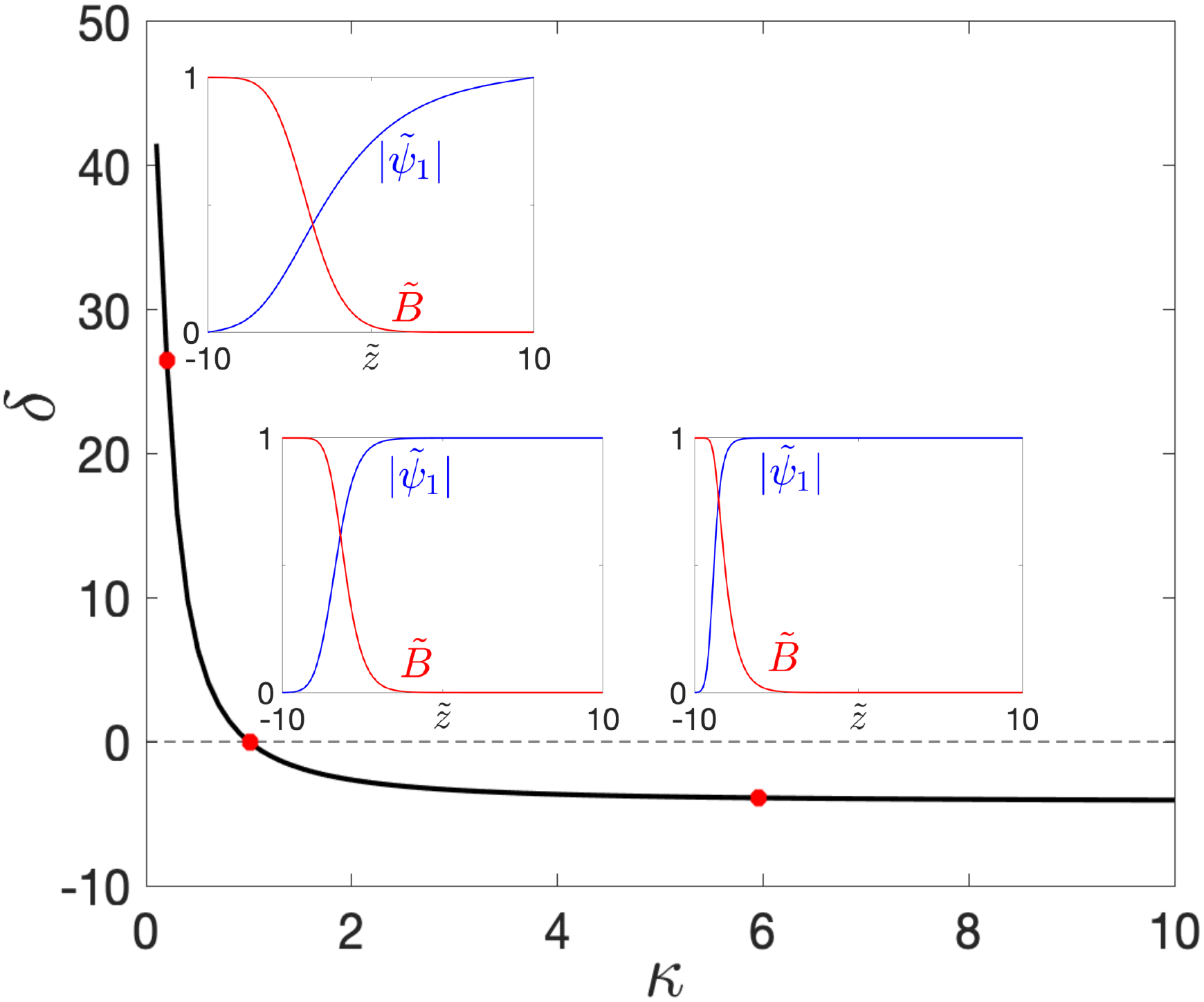}
    \caption{(Color online) Dimensionless surface energy parameter $\delta$ ($=4 \sigma_{\textrm{ns}}/(\delta_0 \beta_1 \abs{\psi_\infty}^4)$) for the three-band chiral system with phase frustration as a function of the Ginzburg-Landau parameter $\kappa$ (black line). The insets show the solutions $|\tilde{\psi}_1|$ ($=|\tilde{\psi}_2|$) and $\tilde{B}$ of the dimensionless Ginzburg-Landau equations for $\kappa = 0.2,~ 1,~ 6$ (red dots). The dashed line $(\delta=\sigma_{\textrm{ns}}=0)$ intersects the curve $\delta(\kappa)$ at $\kappa=1$.}
    \label{fig:surf_en_chiral}
\end{figure}

Using the numerical solutions for $\tilde{\psi}_{1}$, $\tilde{\psi}_{2}$ and $\tilde{B}$ in this expression we obtain the evolution of the surface energy parameter $\delta$ ($=4 \sigma_{\textrm{ns}}/(\delta_0 \beta_1 \abs{\psi_\infty}^4)$) with $\kappa$, shown in Fig.~\ref{fig:surf_en_chiral}. Like for the non-degenerate $N$-band case, the chiral three-band system -- with equal repulsive interband interactions and equal DOS for all the bands -- shows a single transition from $\delta>0$ to $\delta<0$. The sign change occurs at $\kappa=1$, where $\kappa$ is defined in terms of the microscopic parameters, according to Eq.~\eqref{eq:dim_par_chiral}. This redefined critical value for the chiral case, separating the type-I and type-II regimes, emerges as a direct consequence of the degeneracy of the solutions of the gap equation. Hence, we find that the dichotomy between type-I and type-II superconductors, described by a single GL parameter, is preserved for chiral three-band superconductors -- provided their microscopic parameters comply with the symmetries utilized in the model.

\section{Conclusions}

We have explored the surface energy of multiband superconductors within the Ginzburg-Landau framework. We obtained a general formula for an arbitrary number of bands which is fully parameterized by the critical temperature, the band-resolved electronic density of states, and the superconducting gap functions of the different bands. This approach also yielded a general expression for the thermodynamic critical magnetic field of multiband superconductors.

We have subsequently applied this approach to two distinct cases: (i) $N$-band superconductors with only attractive interactions between the bands and non-degenerate solutions to the gap equation, and (ii) a chiral three-band superconductor with phase frustration. We have demonstrated that the Ginzburg-Landau equations can be written in terms of a single Ginzburg-Landau parameter $\kappa$ for both cases. Next, we have numerically solved the Ginzburg-Landau equations for a superconductor-normal interface, to obtain the evolution of the surface energy as a function of $\kappa$. This analysis has demonstrated distinct regimes with positive and negative surface energies for both cases, corresponding to type-I and type-II superconductors respectively. 

Finally, we have applied this approach to several multiband superconductors of prime interest, based on microscopic parameters obtained from first-principles calculations. Our calculations for MgB$_2$ showed a marked reduction of the Ginzburg-Landau parameter $\kappa$ in the monolayer limit. We also calculated the thermodynamic critical magnetic field of metallic hydrogen, demonstrating elevated values as a result of the strong superconducting gaps in this system.  

\begin{appendix}

\section{Numerical methods}
\label{appendix_num_meth}

All numerical computations were performed within MATLAB with double precision. First, the Ginzburg-Landau equations for the non-degenerate $N$-band case (Eq.~\eqref{eq:multibandGL}), accompanied by the boundary conditions (Eq.~\eqref{eq:multibandBC}) define a boundary value problem for $(\tilde{\psi},d\tilde{\psi}/d\tilde{z},\tilde{B},\tilde{A})$ which was solved using the boundary value problem solver \texttt{bvp4c}, which is a fourth-order collocation scheme. Physically, the boundary conditions are defined at $\pm\infty$ and are therefore not numerically tractable, hence we defined them at $\tilde{z} = \pm 10$. Numerical experiments showed that this interval is broad enough to allow the solutions to converge to the boundary values well before reaching the boundary. The number of grid points and their positions are automatically tuned during the execution of the solver, but the initial grid was an equidistant grid with 500 grid points. The same technique was used for the chiral case (Eqs.~\eqref{eq:chirdimless} and \eqref{eq:chiralBC}) to compute $\tilde{A},\tilde{B}$ and the real and imaginary parts of $\tilde{\boldsymbol{\psi}}$. These solutions were used to compute the surface energy by evaluating their integrals (Eqs.~\eqref{eq:SEmulti}, \eqref{eq:SEchir}) by means of the trapezoidal rule. The motivation to choose this quadrature rule was twofold. First it is clear that the solutions to the GL equations for both the non-degenerate multiband and the chiral case are well-behaved and monotonic. The integrand of the surface energy only contains fourth-order powers of the order parameters and second order powers of the magnetic field and therefore is equally well-behaved. This allows the use of a simple and second-order accurate quadrature rule such as the trapezoidal rule.

\section{Implementation of boundary conditions for the chiral case}
\label{appendix_BCs_chiral}

Since we have two order parameters $\tilde{\psi}_1,\tilde{\psi}_2$, we cannot assume both to be real-valued functions. Therefore, we need to split the Ginzburg-Landau equations for $\tilde{\psi}_1,\tilde{\psi}_2,A$ into equations for Re$(\tilde{\psi}_1)$, Re$(\tilde{\psi}_2)$, Im$(\tilde{\psi}_1)$, Im$(\tilde{\psi}_2)$ and $A$. For notational simplicity we write
\begin{align*}
\tilde{\psi}_1 = u_1+i v_1~,~ \tilde{\psi}_2 = u_2 + i v_2~.
\end{align*}
Notice that the left-hand sides of the Ginzburg-Landau equations are linear in $\tilde{\psi}_1,~\tilde{\psi}_2$ and $A$. Therefore taking real and imaginary parts of the equations, we find the following system of differential equations
\begin{align*}
& \frac{d^2u_1}{d\tilde{z}^2}=\kappa^2\left[\frac{1}{4}\left(3(u_1^2+v_1^2)+2(u_2^2+v_2^2)\right)u_1+\frac{1}{4}\left((u_2^2-v_2^2)u_1+2u_2v_2v_1\right)+(-1+\tilde{A}^2)u_1\right]~,\\
& \frac{d^2v_1}{d\tilde{z}^2}=\kappa^2\left[\frac{1}{4}\left(3(u_1^2+v_1^2)+2(u_2^2+v_2^2)\right)v_1+\frac{1}{4}\left(-(u_2^2-v_2^2)v_1+2u_2v_2u_1\right)+(-1+\tilde{A}^2)v_1\right]~,\\
& \frac{d^2u_2}{d\tilde{z}^2}=\kappa^2\left[\frac{1}{4}\left(3(u_2^2+v_2^2)+2(u_1^2+v_1^2)\right)u_2+\frac{1}{4}\left((u_1^2-v_1^2)u_2+2u_1v_1v_2\right)+(-1+\tilde{A}^2)u_2\right]~,\\
& \frac{d^2v_2}{d\tilde{z}^2}=\kappa^2\left[\frac{1}{4}\left(3(u_2^2+v_2^2)+2(u_1^2+v_1^2)\right)v_2+\frac{1}{4}\left(-(u_1^2-v_1^2)v_2+2u_1v_1u_2\right)+(-1+\tilde{A}^2)v_2\right]~,\\
& \frac{d^2\tilde{A}}{d\tilde{z}^2}= \tilde{A}(u_1^2+u_2^2+v_1^2+v_2^2)~,
\end{align*}
accompanied by the following boundary conditions
\begin{align*}
\begin{cases}
    u_1~,~v_1~,~u_2~,~v_2=0~,~\tilde{B}=1~,\qquad  \tilde{z}\to-\infty,\\
    u_1=1~,~v_1=0~,~u_2=\cos(\delta\phi)~,~v_2=\sin(\delta\phi)~,~\tilde{B}=0~, \qquad \tilde{z}\to+\infty~,
\end{cases}
\end{align*}
As stated in Sec.~\ref{sec:chiral}, we have two cases: $\delta\phi = \pi/2+k2\pi$ and $\delta\phi = 3\pi/2+k2\pi,~ k\in\mathbb{N}$. This translates to the following conditions for $\tilde{z}\to +\infty$,
\begin{align*}
\begin{cases}
    u_2=0~,~v_2=1 ~~\qquad\textrm{(case 1)}~,\\
    u_2=0~,~v_2=-1 \qquad\textrm{(case 2)}~.
\end{cases}
\end{align*}
In this form we can compute the solutions numerically because the above problem can easily be translated into a first order boundary value problem.

In MATLAB we computed solutions to the above system of equations for different values of $\kappa$. As for the regular multiband case, we again made an approximation of the boundary conditions at $\tilde{z} = \pm10$.

Finally we rewrite the expression for the surface energy using $u_1,u_2,v_1,v_2$:
\begin{align*}
\delta = \int_{-\infty}^{+\infty} \left[ \frac{-16}{9}(u_2^2+v_2^2)^2-\frac{1}{9}\left((\sqrt{3}u_1+u_2)^2+(\sqrt{3}v_1+v_2)^2\right)^2-\frac{1}{9}\left((\sqrt{3}u_1-u_2)^2+(\sqrt{3}v_1-v_2)^2\right)^2+4(\tilde{B}-1)^2 \right] d\tilde{z}~.
\end{align*}
We performed numerical integration for the above integral using the trapezoidal rule, in the same way as for the non-degenerate multiband case. 

\section{Microscopic parameters}
\label{AppendixC}

\begin{table}[h]
\centering
\begin{tabular}{|c| c| c| c|} 
\hline
band $i$ & $N_{\mathrm{F},i}$ (eV$^{-1}$, per u.c.) & $v_{\mathrm{F},i}$ ($10^7$ cm/s) & $\Delta_{i}(0)$ (meV) \\ 
 \hline
 $\sigma$ & 0.2958 & 5.496 & 7 \\
 \hline
 $\pi$ & 0.4092 & 9.396 & 3 \\ 
 \hline
\end{tabular}
\caption{Partial DOS, Fermi velocities, and average superconducting gap at $T=0$ for the two band condensates of bulk MgB$_2$.}

\label{tab:bulk_mgb2}
\end{table}

\begin{table}[h]
\centering
\begin{tabular}{|c| c| c| c|} 
\hline
band $i$ & $N_{\mathrm{F},i}$ (eV$^{-1}$, per u.c.) & $v_{\mathrm{F},i}$ ($10^7$ cm/s) & $\Delta_{i}(0)$ (meV) \\ 
 \hline
 $\sigma$ & 0.3972 & 6.46 & 3.3 \\
 \hline
 \textit{S} & 0.3769 & 4.23 & 2.7 \\
 \hline
 $\pi$ & 0.1610 & 7.27 & 1.4 \\
 \hline
\end{tabular}
\caption{Partial DOS, Fermi velocities, and average superconducting gap at $T=0$ for the three band condensates of ML MgB$_2$.}
\label{tab:ml_mgb2}
\end{table}

\begin{table}[h]
\centering
\begin{tabular}{|c| c| c| c| c| c|} 
\hline
band $i$ & $N_{\mathrm{F},i}$ (eV$^{-1}$, per u.c.) & $\Delta_{i}(0)$ (meV) \\
 \hline
 1 &  0.4408 & 32 \\
 \hline
 2 & 0.8381 & 25  \\
 \hline
 3 & 6.1607 & 22  \\
 \hline
\end{tabular}
\caption{Partial DOS and fitted average gap functions at $T=0$ for metallic hydrogen \cite{PhysRevLett.100.257001}.}
\label{tab:hydrogen}
\end{table}

The microscopic parameters used in Sections \ref{subsec:metallicH} and \ref{subsec:application_MgB2} are provided in Tables \ref{tab:bulk_mgb2}--\ref{tab:hydrogen} for metallic hydrogen and bulk and ML MgB$_2$, respectively. The Fermi velocities were calculated from the electronic band structures through $\mathbf{v}_{\mathrm{F}} = \hbar^{-1} \boldsymbol{\nabla}_{\textbf{k}}\varepsilon_{\textbf{k}}\mid_{\varepsilon_{\textbf{k}}=E_{\mathrm{F}}}$. The resulting Fermi velocity fields were averaged over the $k$-points for each band separately. The band-resolved DOS values were obtained by integrating the Kohn-Sham eigenvalues belonging to specific bands, using a Gaussian approximation for the Dirac delta function. 

The electron-phonon coupling matrix of bulk MgB$_2$ was measured to consist of $\lambda_{\sigma \sigma}=0.84$, $\lambda_{\sigma \pi}=0.19$ and $\lambda_{\pi \pi}=0.39$ \cite{Kuzmichev2014}. For ML MgB$_2$, the electron-phonon interaction matrix, decomposed into contributions from scattering of electrons from band $i$ to band $j$, was obtained via \cite{PhysRevLett.87.087005}
\begin{align*}
G_{ij}=2\left(N_{\mathrm{F},i}N_{\mathrm{F},j}\right)^{-1}\sum_{\textbf{k}\textbf{q}\nu} \omega_{\textbf{q}\nu}^{-1} \left|g_{\textbf{k}i,\textbf{k}+\textbf{q}j}^{\nu}\right|^2 \delta \left(\varepsilon_{\textbf{k}i} \right) \delta \left(\varepsilon_{\textbf{k}+\textbf{q}j} \right)~,
\label{Eq:G_matrix}
\end{align*}
where $\omega_{\textbf{q}\nu}$ is the phonon dispersion for phonon branch $\nu$ at wave vector \textbf{q}, $\varepsilon_{\textbf{k}i}$ the electronic dispersion for band $i$ at wave vector \textbf{k}, and $g_{\textbf{k}i,\textbf{k}+\textbf{q}j}^{\nu}$ the electron-phonon coupling matrix elements (using DFPT results from Ref.~\citenum{PhysRevB.96.094510}). This yields for ML MgB$_2$:
\begin{align*}
G=\begin{pmatrix}
g_{\sigma \sigma} & g_{\sigma S} & g_{\sigma \pi}\\
g_{S \sigma} & g_{S S} & g_{S \pi}\\
g_{\pi \sigma} & g_{\pi S} & g_{\pi \pi}
\end{pmatrix}=\begin{pmatrix}
0.115 & 0.075 & 0.050\\
0.075 & 0.047 & 0.014 \\
0.050 & 0.014 & 0.038\\
\end{pmatrix}
\end{align*}
in units of Ha $\times ~ V_{\mathrm{uc}}$ (the latter being the unit cell volume). The unit cell volume of bulk MgB$_2$ was obtained as $29.0642 \cdot 10^{-24}$ cm$^3$ from our DFT calculations. The unit cell volume of the bulk structure was also used as characteristic unit cell volume for the ML case.

The $\Delta_{i}(0)$ values in Table \ref{tab:hydrogen} were fitted so as to reproduce the gap values obtained from SCDFT calculations in Ref.~\citenum{PhysRevLett.100.257001} in the range $\left[0.7 - 1 \right]  T_{\mathrm{c}}$ according to the GL relation
\begin{align*}
\Delta_i^{(0)}(T)=\Delta_{i}(0) \sqrt{\tau} = \Delta_{i}(0) \sqrt{1-\frac{T}{T_{\mathrm{c}}}}~. 
\end{align*}
The unit cell volume of metallic hydrogen under a pressure of 414 GPa is $4.3359 \cdot 10^{-24}$ cm$^3$ \cite{PhysRevB.81.134505}. 
\end{appendix}

\begin{acknowledgments}
\noindent J.B. is a senior postdoctoral fellow of Research Foundation-Flanders (FWO). The computational resources and services were provided by the VSC (Flemish Supercomputer Center), funded by the FWO and the Flemish Government -- department EWI. We thank Gianni Profeta from the University of L'Aquila (Italy) for providing additional data on metallic hydrogen from Ref.~\citenum{PhysRevLett.100.257001}. We also acknowledge the KU Leuven (Belgium) for giving L.B. the opportunity to contribute to the research presented here within the framework of the Honours Programme, under the supervision of J.B.
\end{acknowledgments}

\bibliography{Refs}

\end{document}